\documentclass[12pt,titlepage]{article}
\usepackage{epsfig}
\usepackage{lscape}
\usepackage{amsmath,epsf,color,colordvi,shadow,pifont}

\begin{document}

\begin{titlepage}
\begin{center}
{\LARGE {\bf  Vacuum energy, holography and a quantum portrait of the visible 
Universe}}
 \\
\vskip 1cm

{\large P. Bin\'etruy } \\
binetruy@apc.univ-paris7.fr \\
{\em AstroParticule et Cosmologie, 
Universit\'e Paris Diderot, CNRS, CEA, Observatoire de Paris, Sorbonne Paris 
Cit\'e} \\
{B\^atiment Condorcet, 10, rue Alice Domon et L\'eonie Duquet,\\
F-75205 Paris Cedex 13, France}
\end{center}

\centerline{ {\bf Abstract}}

\indent 
Describing the presently observable Universe as a self-sustained condensate of 
gravitons of size $H_0^{-1}$, with large occupation number $N$, we argue that 
the most probable value for the quantum vacuum energy is of the order of
the critical energy density, as observed.   

\vfill  
\end{titlepage}
\def\noi{\noindent}
\def\sq{\hbox {\rlap{$\sqcap$}$\sqcup$}}
\def\1{{\rm 1\mskip-4.5mu l} }

\newpage
\pagenumbering{arabic}
\section{Computing vacuum energy in the context of gravity} 
\label{sect:1}
Vacuum energy, which we define loosely as the energy of the fundamental state, 
is a measurable quantity only in the context of gravity. Indeed, in a 
non-gravitational context, only differences of energy, such as forces, can 
be mesured (a well-known example is the Casimir effect). On the other hand, the 
presence of the energy-momentum tensor on the right-hand side of Einstein's 
equations shows that any form of energy impacts on the geometry of spacetime. 
In particular, the Friedmann equation which can be deduced expresses the fact 
that any form of energy participates to the expansion of the Universe. Hence 
energy, in particular the energy of the vacuum, can be measured absolutely.

We know from observation that space is flat; hence the energy density of the 
vacuum $\rho_{vac}$ is smaller than the critical density:
\begin{equation}
\label{eq:1}
\rho_{vac} < \rho_c \equiv {3 H^2_0 \over 8 \pi G_N} \ .
\end{equation}
This should be compared with a naive guess estimate: in the context of a 
quantum theory of gravity, one expects on dimensional grounds that $\rho_{vac}$
scales as $m_P/l_P^3$, where $m_P$ is the (reduced) Planck mass scale of 
quantum gravity i.e. $m_P \equiv \sqrt{\hbar c/(8\pi G_N)}$  and 
$l_P \equiv \hbar / (m_P c)$ is the Planck length (from now on, we set $k=
c = 1$).
In other words,
\begin{equation}
\label{eq:2}
\rho_{vac} \sim {m_P \over l_P^3}  = {1 \over \hbar(8 \pi G_N)^2} \ .
\end{equation}
This however does not take into account the specificity of gravity. Let us 
consider a spherical region of radius $R$. It cannot contain more mass-energy 
than a black hole of same size i.e. of Schwarzschild radius $R_S = R$. Hence, 
using $R_S = 2 G_N M$ ($M$ mass of black hole), the energy 
$E$ in this spherical region satisfies the relation $E < M = R/(2 G_N)$  and, 
disregarding constants of order one,
\begin{equation}
\label{eq:3}
\rho_{vac} = E/(4 \pi R^3/3) < {3 \over 8\pi G_N R^2} \ .
\end{equation}
If we extend this to the whole observable Universe of radius $H_0^{-1}$, this 
gives
\begin{equation}
\label{eq:4}
\rho_{vac} < {3H_0^2 \over 8 \pi G_N} \ .
\end{equation}  
Let us note the surprising similarity with the observational constraint 
(\ref{eq:1}).

The rationale behind the limit (\ref{eq:3}) is the fact that, from a 
gravitational point of view, each Planck cell of size $l_P = \hbar / m_P$
within a  volume $R^3$ cannot host a maximal energy $m_P$: this would lead 
to a total mass $R^3/l_P^{3} \times m_P$ or $M_{BH}(R/l_P)^2$, where $M_{BH}
\sim Rm_P^2/\hbar$ is the
mass of a black hole with size $R$, and thus to gravitational 
collapse whenever we consider a volume element larger than an elementary scale
(i.e. $R > l_P$). Hence gravitational collapse prevents the ultraviolet 
cut-off of the quantum theory to reach its maximal value $m_P$ in a large 
fraction of elementary shells. At the level of 
the 
whole observable universe, this provides a connection between the microscopic 
ultraviolet scale $m_P$ and the cosmological infrared scale $H_0^{-1}$ which is 
expressed as the bound (\ref{eq:3}).

Obviously, these ideas are reminiscent of the ones associated with holography 
and entropy bounds \cite{'tHooft:1993gx,Susskind:1994vu} 
(see \cite{Bousso:2002ju} and references therein), as applied to cosmology
\cite{Banks:1995uh,Horava:1997dd,Fischler:1998st,Easther:1999gk,Veneziano:1999ts,Kaloper:1999tt}. Indeed, the largest 
number of degrees of freedom that we can
pack in a region of size $R$ is the one corresponding to a black hole of 
size $R$, i.e. $N < (R/l_P)^2$ since the degrees of freedom of a black 
hole lie on the horizon surface. Since for each individual cell the Poissonian 
fluctuation in energy is $\Delta \epsilon \sim m_P$, the energy for the overall 
fluctuations is $\Delta E^2 = N m_P^2$, which corresponds to an energy density
\begin{equation}
\label{eq:5}
\rho_{vac} = {\Delta E \over R^3} = {\sqrt{N} m_P \over  R^3} 
< {m_P^2 \over \hbar R^2}  \ ,  
\end{equation}
again consistent with (\ref{eq:3}).

Since the upper value corresponds to the maximal entropy, one expects on 
statistical grounds (see next section) that
\begin{equation}
\label{eq:5bis}
\rho_{vac} =  {m_P^2 \over \hbar R^2}  \ ,  
\end{equation}
which would yield when applied to the whole observable Universe ($R=H_0^{-1}$)
the well known relation \cite{Cohen:1998zx,Hsu:2004jt}
\begin{equation}
\label{eq:5ter}
\rho_{vac} =  { \hbar  \over l_P^2 H_0^{-2}}  \ .  
\end{equation}

\section{A quantum portait of the visible Universe}
   
The preceding ideas regarding vacuum energy are following the same lines as the 
discussion of gravity theory as a classicalized theory. We will see that there 
are indeed some strong similarities between some of the concepts developped 
within classicalized gravity and a possible description of our observable 
universe.  

Let us consider our observable Universe. It consists approximately of 30\% of 
matter and 70\% of dark energy. We will make the hypothesis that dark energy is 
vacuum energy and will neglect for the time being the subdominant matter 
component: we attempt to describe first a Universe with only vacuum energy.

This Universe is obviously a classical object, the most classical of all in 
some sense but, because fundamental forces are described by quantum physics, 
it should have as well a quantum description: a graviton bound-state 
with a very high occupation number $N \gg 1$.

Let us consider first an individual quantum state of graviton with energy 
$\epsilon \sim 
\hbar k \sim \hbar H_0$ (since $k$ is typically the inverse of the Hubble 
horizon length $H_0^{-1}$). Then the total energy $E$ in the observable universe
is simply $N\epsilon$ where $N$ is the total occupation number. Now 
we have seen in the preceding section that $E<H_0^{-1}/(2G_N)$, hence
\begin{equation}
\label{eq:6}
N = {E \over \epsilon} < {1 \over G_N \hbar H_0^2}={1 \over l_P^2 H_0^2} \ ,
\end{equation}
where we used $l_P^2 = \hbar G_N$. The limit (\ref{eq:6}) represents the highest
possible value for $N$; in other words, maximal classicality of the observable 
universe is reached for 
\begin{equation}
\label{eq:7}
N = (l_P H_0)^{-2} \ .
\end{equation}

This is reminiscent of the quantum N-portrait of a black hole as described by
Dvali and Gomez \cite{Dvali:2011aa,Dvali:2012gb,Dvali:2012en}. This is 
obviously not surprising since
the black hole represents the most classical object within a region of a given 
size. We will pursue the analogy and describe some of the properties of the 
universe seen as a Bose-Einstein condensate of $N = (l_P H_0)^{-2}$  soft 
gravitons of wavelength $\lambda \sim H_0^{-1}$ which are weakly interacting 
(their dimensionless coupling is $\hbar G_N / \lambda^2 \sim (l_P H_0)^2$). 

First, as emphasized in \cite{Dvali:2011aa}, the condensate is self-sustained 
only if its size $H_0^{-1}$ does not overcome too much its Schwarzschild radius
$R_S$ otherwise not only the gravitons are extremely weakly coupling to one 
another but also the interaction of one graviton with the collective 
gravitational energy is negligible. This  suggests that our own observable 
universe saturates the bound (\ref{eq:6}) and thus satisfies
$N = (l_P H_0)^{-2}$, i.e. it is maximally classical. 

Second, just as in the case of a black hole \cite{Dvali:2011aa}, there is a 
`thermal spectrum of temperature
\begin{equation}
\label{eq:7}
T = {\hbar \over \sqrt{N} l_P} = \hbar H_0 \ .
\end{equation}
This should be compared to the famous result according to which an observer
in de Sitter space (with Hubble parameter $H$) feels as if he is in a thermal 
bath of temperature $T = \hbar H/(2\pi)$. The result (\ref{eq:7}) is thus 
consistent with the fact that, if the vacuum energy density is dominant, we are 
in a de Sitter phase. 

Just as a black hole evaporates, one expects that the whole observable universe 
will decay after a time:
\begin{equation}
\label{eq:8}
t_{dec} = N^{3/2} l_P = l_P^{-2} H_0^{-3} = N H_0^{-1} \ ,
\end{equation}
which is thus much larger than the present age of the Universe\footnote{This is 
to be contrasted with the result obtained with dark energy models where the
Universe collapses within a time of the order of $H_0^{-1}$ 
\cite{Kallosh:2003mt}.}. 

Finally, one may define the entropy of the visible Universe as 
\cite{Dvali:2011aa}
\begin{equation}
\label{eq:9}
S = N 
\end{equation}
Following a Boltzmann distribution, we expect  that the probability for $N$ to 
have a value in the interval between $N$ and $N+dN$ will be given by:
\begin{equation}
\label{eq:10}
w(N) dN = \hbox{cst}\  e^S = \hbox{cst} \ e^N \ .
\end{equation}
The probability is maximum for the largest possible value of $N$ compatible 
with the limit (\ref{eq:6}), that is for  $N = (l_P H_0)^{-2}$ which corresponds
to 
\begin{equation}
\label{eq:11}
\rho_{vac} = {N \epsilon \over (H_0^{-1})^3} = {\hbar \over l_P^2 H_0^{-2}} \ .
\end{equation}
as in (\ref{eq:5ter}). 

We note that , by writing
\begin{equation}
\label{eq:12}
e^S = e^N = e^{l_P^{-2} H_0^{-2}} = e^{\hbar/(\rho_{vac}l_P^4)} = e^{1/(\Lambda l_P^2)} \ ,
\end{equation}
where we introduced the cosmological constant $\Lambda = 8\pi G_N \rho_{vac}$, 
we recover the distribution $w(\Lambda)d\Lambda = \hbox{cst} \ e^{1/(\Lambda 
l_P^2)}$ proposed by Hora\v{v}a and Minic \cite{Horava:2000tb}. However, while 
these authors inferred from such a distribution that a vanishing cosmological 
constant $\Lambda$ has maximal probability, we draw a different conclusion: 
taking $l_P$ as given by the theory and $H_0$ as imposed by observation, we 
infer that the maximal probability corresponds to maximal $N$ and thus to
$\rho_{vac}$ given by (\ref{eq:11}).    

\section{The cosmological evolution of our Universe}

The cosmological scenario that emerges from the preceding considerations is 
both familiar and very different from what is usually described. We assume that
the Universe emerges from the quantum epoch (characterized by a length scale 
$l_P$) in a quantum state which has many 
classical realizations i.e. which can be projected onto many different 
classical states. Projection occurs through the measurement process, that 
identifies a (visible) Universe of size $H_0^{-1}$. In the absence of matter, 
this Universe is a classical condensate of weakly interacting gravitons. Its 
stability imposes the condition (\ref{eq:4}) on the vacuum energy density. 
Moreover, the upper limit $\rho_{vac} = \hbar/(l_P H_0^{-1})^2$ corresponds to
maximal probability and maximal classicality. 

It should be stressed that the relation $\rho_{vac} = \hbar/(l_P H_0^{-1})^2$
is only valid at the time of measurement i.e. at the time where the state of 
the Universe is projected onto the classical state. The classical Universe
then deploys itself in time (backward and forward) in the standard way.
In particular, in the case where matter and radiation are negligible, the 
Universe falls in  de Sitter expansion as we have discussed above (with 
associated radiation at temperature $T \sim \hbar H_0$). The classical 
description has obviously a limited range of validity, namely $kT \ll m_P$.

A useful analogy is provided by the simple double slit interference experiment 
with electrons. If one is interested in the time of flight of the electrons from
the source to the screen, one may follow step by step the motion of the
electron (and thus identify which slit it went through): this time 
of flight is a classical quantity and can thus be measured classically. On the 
other hand, one will be losing the interference pattern on the screen. If one 
wants to recover the interference pattern, one should avoid tracing the 
electron through its evolution. Similarly, if one wants to understand the  
amount of vacuum energy (a quantum observable), one should not trace the
Universe through its evolution. Instead, one may compute probabilities for 
measuring a given value, once we observe the Universe (that is, now). Other 
aspects of cosmology which are purely classical (from the point of view of 
gravity) may reliably be computed by following the evolution of the Universe.
  
This should be contrasted with cosmologies proposed along similar lines which 
assumed the relation  $\rho_{vac} \sim \hbar/(l_P H^{-1})^2$ throughout the 
evolution of the Universe. It is easily seen that they cannot describe
dark energy \cite{Hsu:2004ri}\footnote{Indeed, if $\rho_{vac} = \alpha 
H^2/(8\pi G_N)$, with $\alpha$ un unknown constant, then the Friedmann 
equation $H^2 = (8\pi G_N/3)(\rho_{vac} + \rho)$ may be rewritten as 
$H^2=8 \pi G_N \rho /(3-\alpha)$, which amounts to a mere rescaling of Newton's 
constant.}. 

One may now add quantum fields to describe matter, radiation, inflaton into our 
description of the Universe\footnote{One could imagine applying the preceding 
ideas not just to vacuum energy but to the total energy density $\rho_T$. In 
this case, one would reach the conclusion that $\rho_T = 3H_0^2/(8\pi G_N)$ 
i.e. that the Universe is spatially flat. But again, our argument only applies 
to the quantum observable that is the vacuum energy. There remains the 
possibility of bosonic dark matter participating to the energy budget (just as 
bosons may account for the baryonic component in a black hole \cite{Dvali:2012rt}).}. 
By doing this, one may 
wonder whether past phase transitions or inflationary epochs may lead to a 
change in the vacuum energy. But again, in the framework presented, the vacuum
energy has the value it has because the observed Universe has the size it has
($H_0^{-1}$).
An inflation scenario is still needed in order to explain the flatness of the 
Universe but the true (present) ground state of the inflaton field must 
correspond to $\rho_{vac} \sim \hbar/(l_P H_0^{-1})^2$ in order to comply with
the observation that the Universe is large (of size $H_0^{-1}$).

\vskip .5cm
To conclude, we propose to identify the presently observable universe to the 
same Bose-Einstein condensate of gravitons that describes a black hole. This 
seems at first to contradict our view of a black  hole as a very dense object, 
but one should remember that the density of a black hole decreases as the 
inverse square of its radius. Indeed, the density of the presently observable 
universe (of radius $H_0^{-1}$)
has the right order of magnitude. Moreover, it appears plausible that the 
Universe, when we observe it, is a self-sustainable condensate of gravitons
with a classical behaviour. This is exactly what is a black hole, only at a 
different length scale.

This allows us to understand the order of magnitude of the vacuum energy 
density, in agreement with observation. The value obtained is such because the 
observed universe is large. This provides a new twist to the question ``Why
does vacuum energy become dominant now?'' and correspondingly a different 
solution to this problem.

We focused in this paper on the main component of the Universe i.e. dark energy
(which, in our case, is vacuum energy). This departs from the standard attitude
which, for historical reasons, considers dark energy as an ``extra'' component. 
To us, it appears that one should first explain the 
dark universe before addressing the question of luminous matter, which appears 
to be a detail (though an important one) in the present Universe. When one 
tries to
add matter to the proposed scheme, new and interesting possibilities arise, in 
particular for dark matter. 
   
Let us conclude by stressing again the main difference with respect to the 
standard cosmological scenarios. Measurement (in the quantum mechanical sense)
realizes the Universe as we know it, among the many possible classical 
realizations. The type of graviton condensate that forms our {\em classical} 
spacetime allows to identify continuous time and space, and to describe 
classically the Universe and its evolution (backward and forward in time). 
Obviously, the description is valid only for $kT \ll m_P$. In a certain sense there 
is no big bang but a multiplicity of different potential classical universes.

{\bf Acknowlegments:} I wish to thank Alexis Helou, who worked for internship  
on the topics covered by this article, for useful discussions and computations. 
I also thank Gia Dvali for discussion.

\end{document}